\documentclass[twocolumn,showpacs,aps]{revtex4}

\usepackage{graphicx}% Include figure files
\usepackage{dcolumn}% Align table columns on decimal point
\usepackage{bm}% bold math

\begin{document}
\title {Manipulating vortex motion by thermal and Lorentz force in high
temperature superconductors}
\author {Z. Wang}
\author{L. Shan}
\author{Y. Z. Zhang}
\author{J. Yan}
\author{F. Zhou}
\author{J. W. Xiong}
\author{W. X. Ti}
\author{H. H. Wen\footnote{corresponding author. Electronic address:
hhwen@aphy.iphy.ac.cn}}
%\author{H. H. Wen\corauthref{cor}}
%\corauth[cor]{Corresponding author.}
%\ead{hhwen@aphy.iphy.ac.cn}
\affiliation{National Laboratory for superconductivity, Institute
of Physics,  Chinese Academy of Sciences, Beijing National
Laboratory for Condensed Matter Physics, P. O. Box 603, Beijing,
100080, China}

\begin{abstract}
By using thermal and Lorentz force, the vortex motion is
successfully manipulated in the mixed state of underdoped La$_{2 -
x}$Sr$_{x}$CuO$_{4}$ single crystals and optimally doped
YBa$_{2}$Cu$_{3}$O$_{7 - \delta }$ thin films. A conclusion is
drawn that the strong Nernst signal above $T_{c}$ is induced by
vortex motion. In the normal state, in order to reduce the
dissipative contribution from the quasiparticle scattering and
enhance the signal due to the possible vortex motion, a new
measurement configuration is proposed. It is found that the
in-plane Nernst signal ($H$ $\vert \vert $ $c$) can be measurable
up to a high temperature in the pseudogap region, while the
Abrikosov flux flow dissipation can only be measured up to
$T_{c}$. This may point to different vortices below and above
$T_{c}$ if we attribute the strong Nernst signal in the pseudogap
region to the vortex motion. Below $T_{c}$ the dissipation is
induced by the motion of the Abrikosov vortices. Above $T_{c}$ the
dissipation may be caused by the motion of the spontaneously
generated unbinded vortex-antivortex pairs.

\end{abstract}

\maketitle

\section{Introduction}
One of the core issues in high temperature superconductor is the
origin of the pseudogap above $T_{c}$ in the underdoped region. In
order to understand the physics of the pseudogap, many theoretical
models have been proposed such as spin fluctuation\cite{Pines},
preformed cooper pair\cite{Emery1,Millis}, charge
stripes\cite{Emery2}, \textit{d}-density wave
(DDW)\cite{Affleck,Chakravarty}, etc. Among many of them, the
pseudogap state has been considered as the precursor to the
superconducting state. In the pseudogap region above $T_{c}$, a
significant in-plane Nernst signals has been discovered by Xu
\textit{et al}.\cite{Xu} in the underdoped La$_{2 -
x}$Sr$_{x}$CuO$_{4}$ single crystal, and they attributed this
signal to vortex-like excitations above $T_{c}$. This result has
been confirmed in other families of cuprate superconductors
\cite{WYY1,WYY2}. By doing experiments with the magnetic field
applied along different directions, Wen \textit{et al}.\cite{Wen}
gave a strong indication for a 2D feature of the Nernst effect in
the pseudogap region of underdoped cuprate superconductors. About
the origin of the strong Nernst signal above $T_{c}$, it remains
highly controversial. Wang \textit{et al}.
\cite{WYY2,WYY3}suggested that the large Nernst signal supports
the scenario\cite{Emery1} where the superconducting order
parameter disappears at a much higher temperature instead of
$T_{c}$. Kontani\cite{Kontani} suggested that the pseudogap
phenomena including the strong Nernst signal can be well described
in terms of the Fermi liquid with antiferromagnetic and
superconducting fluctuations. Ussishkin \textit{et
al}.\cite{Ussiahkin} proposed that the Gaussian superconducting
fluctuations can sufficiently explain the Nernst signal in the
optimally doped and overdoped region, but in the underdoped region
the actual $T_{c}$ is suppressed from meanfield temperature by
non-Gassian fluctuation. Tan \textit{et al}.\cite{Tan} believed
that a preformed-pair alternative to the vortex scenario can lead
to a strong Nernst signal. Alexandrov and Zavaritaky\cite{AZ}
proposed a model based on normal state carrier without
superconducting fluctuation, which, as they asserted, `can
describe the anomalous Nernst signal in high-$T_{c}$
cuprate'\cite{AZ}.

It is thus strongly desired to investigate the feature of Nernst
signal below and above $T_{c}$. One possibility is that the strong
Nernst signal above $T_{c}$ is originated from or partly from the
motion of vortex, but the structure and the feature of the
vortices above $T_{c}$ is different from that below $T_{c}$. We
thus measured the in-plane Nernst voltage and the resistance of
the underdoped La$_{2 - x}$Sr$_{x}$CuO$_{4}$ single crystals and
optimally doped YBa$_{2}$Cu$_{3}$O$_{7 - \delta }$ thin films in
the magnetic field perpendicular to the \textit{ab}-plane. Besides
the temperature gradient, we applied a transverse current to
manipulate the possible motion of the vortices. Especially we
measured the longitudinal voltage in a new configuration which may
reduce the dissipative contribution from the quasiparticle
scattering and enhance the signal due to the possible vortex
motion. In this new configuration, a strong signal measured from
the Nernst leads above $T_{c}$ is observed in underdoped samples.
The signal maybe caused by the motion of quasiparticles and
vortex. Our results indicate that the Nernst signal above $T_{c}$
is contributed by vortex. The results were analyzed and we think
that if strong Nernst signal in the pseudogap region is
contributed by the vortex motion, vortices are different below and
above $T_{c}$. Below $T_{c}$ the dissipation is induced by the
motion of the Abrikosov vortices. Above $T_{c}$ the dissipation
may be caused by the motion of the spontaneously generated
unbinded vortex-antivortex pairs (vortex plasma).

\section{Experimental techniques}

\begin{figure}[t]
\includegraphics
[width= .86\columnwidth] {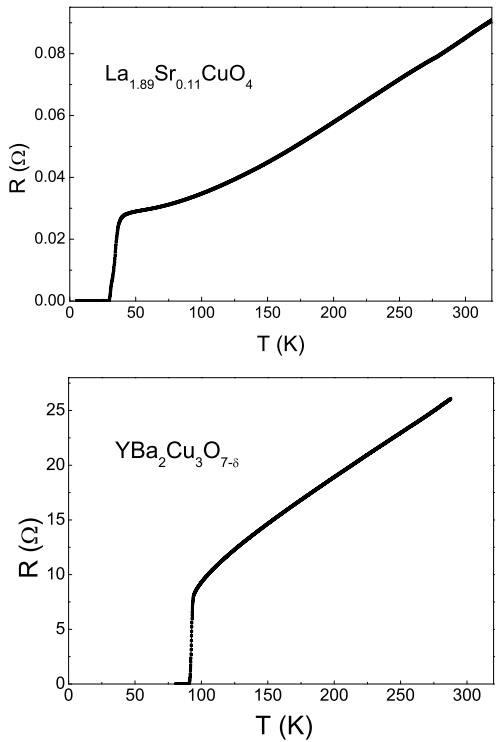} \caption{Temperature dependence of
resistance for La$_{1.89}$Sr$_{0.11}$CuO$_{4}$ single crystal
(upper) and YBa$_{2}$Cu$_{3}$O$_{7 - \delta }$ thin film (lower).}
\label{f1}
\end{figure}

The La$_{1.89}$Sr$_{0.11}$CuO$_{4}$ single crystals measured in
this work were prepared by the traveling solvent floating-zone
technique\cite{Zhou}. The perfect crystallinity of the single
crystals were characterized by X-ray diffraction
patterns\cite{Zhou}. The single crystal sample is shaped into a
bar with the dimensions of of 4.2 mm (length) $\times $ 1 mm
(width) $\times $ 0.5 mm (thickness). The crystal along the length
direction is [110]. The optimally doped
YBa$_{2}$Cu$_{3}$O$_{7-\delta }$ thin films were prepared by
co-evaporation on MgO substrates with dimensions of 10 mm (length)
$\times $ 1 mm (width), and the thickness of the film is about
5000 angstrom. We annealled the YBa$_{2}$Cu$_{3}$O$_{7 - \delta }$
film to underdoping state and measured its Nernst voltage in the
mixed and normal state. The resistance characteristics of the
samples measured is shown in FIG \ref{f1}.

%%%%%%%%%%%%%%%
%%%%%%%%%%%%%%%

\begin{figure}[t]
\includegraphics
[width= .96\columnwidth] {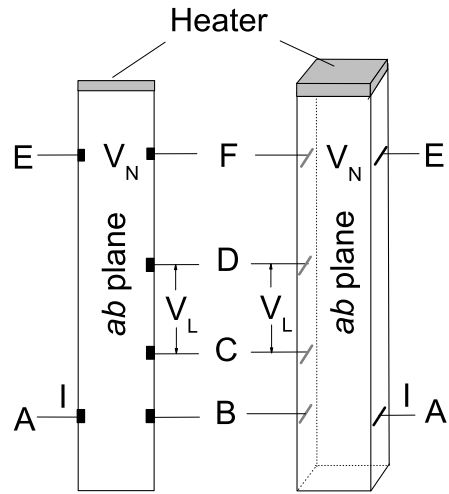} \caption{The configuration for
measuring the Nernst and Longitudinal voltage of
La$_{1.89}$Sr$_{0.11}$CuO$_{4}$ single crystal (right) and
YBa$_{2}$Cu$_{3}$O$_{7 - \delta }$ thin film (left). The Nernst
voltages $V_{N}$ can be obtained between E and F when the samples
are heated by the heaters on one end and the longitudinal voltage
$V_{L}$ can be obtained between C and D when a DC current $I$ is
applied along A-B direction.} \label{f2}
\end{figure}

The measurement configuration in our experiment is illustrated in
FIG. \ref{f2}. The magnetic field is applied along the
\textit{c}-axis of the samples. The $R(H)$ curves of the samples
were obtained by standard four-point method at different
temperatures. A heater with a power of 1 mW (for the single
crystal) or 9 mW (for the thin film) producing the thermal
gradient is adhered fast on one end of the YBa$_{2}$Cu$_{3}$O$_{7
- \delta }$ thin film sample (left) and
La$_{1.89}$Sr$_{0.11}$CuO$_{4}$ single crystal sample (right). The
other end of the samples is adhered fast on a cold sink. Along the
direction E-F (perpendicular to the thermal gradient) the Nernst
voltage signal $V_{N}$ is measured. Two thermometers are attached
onto the sample to detect the temperature gradient of the samples.
In order to control the vortex motion by Lorentz force, we applied
a direct transverse current of 3 mA along A-B and measured the
longitudinal voltage between the points C and D. Since the DC
current here can exert a Lorentz force to the vortices down or
againt the thermal stream direction, the motion of the vortex can
be manipulated by thermal and/or Lorentz force. All leads are
stuck onto the samples by solidified silver paste at corresponding
electrodes with the contact resistance below 0.1 $\Omega$. All
measurements are based on an Oxford cryogenic system ( Maglab-12 )
with temperature fluctuation less than 0.04{\%} and magnetic
fields up to 12 T. The Nernst voltage and the longitudinal voltage
are measured by a Keithley 182-Nanovoltmeter with a resolution of
about 5 nV in present case. During the measurement for Nernst and
longitudinal voltage the magnetic field is applied parallel to
\textit{c} - axis and swept between 12 to -12 T. The Nernst signal
$V_{N}$ is obtained by subtracting the positive field value with
the negative one in order to remove the Faraday signal during
sweeping the field and the possible thermal electric power due to
asymmetry of the electrodes. The Longitudinal signal $V_{L}$ is
obtained by averaging the data obtained from positive field and
the negative field.

%%%%%%%%%%%%%%%%%%%%
%%%%%%%%%%%%%%%%%%%%

\section{Results and discussion}

\subsection{Pure Nernst Signal}

\begin{figure}
\includegraphics
[width= .99\columnwidth] {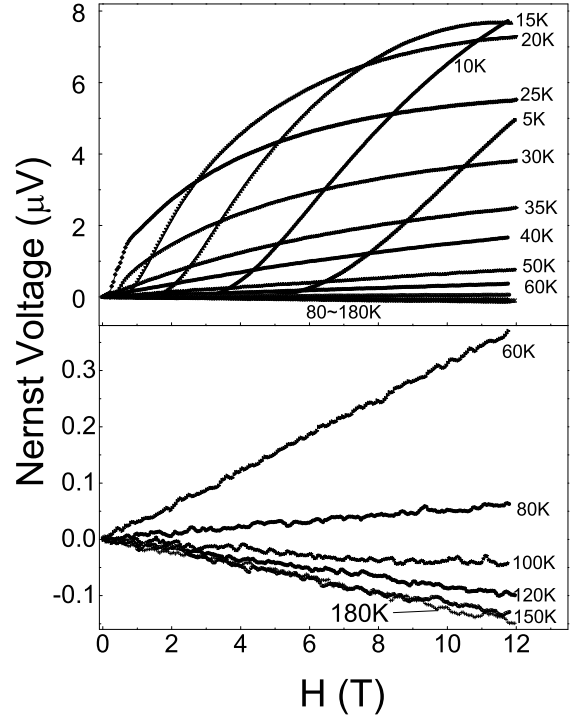} \caption{\label{f3} Nernst voltage
($I$ = 0) of La$_{1.89}$Sr$_{0.11}$CuO$_{4}$ single crystal at
temperatures from 5 K to 180 K (upper panel). In order to make the
result more obvious, the data in the high temperature region (from
60 K to 180 K) are plotted in the lower panel. At temperatures of
5, 10 and 15 K, the vortices are frozen showing a background with
precisely zero signal under relatively low field. A strong
in-plane Nernst signal can be seen in the curves for 20, 25, 30 K,
and can be measured far above $T_{c}$. When the temperature is
above 80 K, the signal becomes negative and gradually approaches a
background.When the temperature is above 150 K the Nernst signal
becomes insensitive to temperature and does not change anymore
with $T$.}
\end{figure}

FIG. \ref{f3}. shows the Nernst voltage measured on the
La$_{1.89}$Sr$_{0.11}$CuO$_{4}$ single crystal with a thermal
gradient along [110] (length) direction at temperatures from 5 K
to 180 K. In low temperature region, the Nernst signal is
dominated by the motion of Abrikosov vortices. One can see that
the Nernst signal is precisely zero when the vortices are frozen
in the case of 5, 10 and 15 K. The flow of vortices after melting
leads to a Nernst signal increasing drastically with $H$. A strong
in-plane Nernst signal resulting from vortices flow can also be
seen in the curves for 20, 25, 30 K, and can be measured far above
$T_{c}$ (29.3 K for our sample). When the temperature is above 80
K, the signal becomes negative and gradually approaches a
background. When the temperature is above 150 K the Nernst signal
becomes insensitive to temperature and does not change anymore
with $T$. Similar results appear in the Nernst voltage measured on
the underdoped YBa$_{2}$Cu$_{3}$O$_{7 - \delta }$ thin film,
though it is not as strong as that of
La$_{1.89}$Sr$_{0.11}$CuO$_{4}$ single crystal sample. But no
strong Nernst signal above $T_{c}$ is observed in optimally doped
YBa$_{2}$Cu$_{3}$O$_{7 - \delta }$ thin film.

%%%%%%%%%%%%%%%%%%%%
%%%%%%%%%%%%%%%%%%%

\subsection{Flux Flow Resistance}

%%%%%%%%%%%%%%%%%%%%
\begin{figure}
\includegraphics
[width= .99\columnwidth] {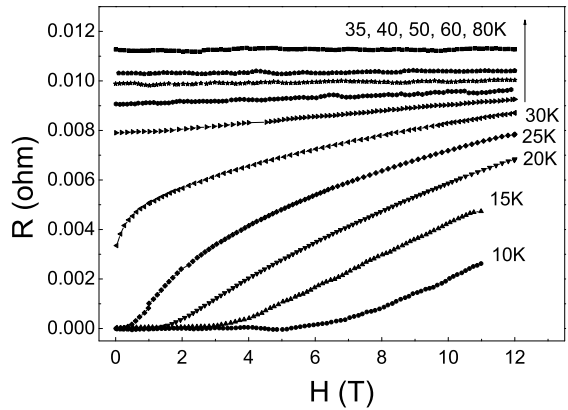} \caption{\label{f4} $H$ dependence
of the resistance of La$_{1.89}$Sr$_{0.11}$CuO$_{4}$ single
crystal at different fixed temperatures without thermal gradient.
The Abrikosov flux flow dissipation can be seen below $T_{c}$, but
the resistance is insensitive to $H$ above $T_{c}$.}
\end{figure}
%%%%%%%%%%%%%%%%%%%

FIG. \ref{f4} shows the $R(H)$ curves of the
La$_{1.89}$Sr$_{0.11}$CuO$_{4}$ single crystal at temperatures
from 10 to 80 K measured by standard four-point method. A zero
resistance can be seen when the flux lattice is frozen at the low
temperature of 10, 15 and 20 K. The Abrikosov flux flow
dissipation after vortex melting can be seen in higher temperature
region. But the curvature due to the motion of Abrikosov vortices
disappears when the temperature is above $T_{c}$. The dissipation
becomes very weakly dependent on $H$ in the normal state, which
can not be explained by the motion of Abrikosov vortices.

\subsection{Longitudinal Signal}

\begin{figure}[b]
\includegraphics
[width= .99\columnwidth] {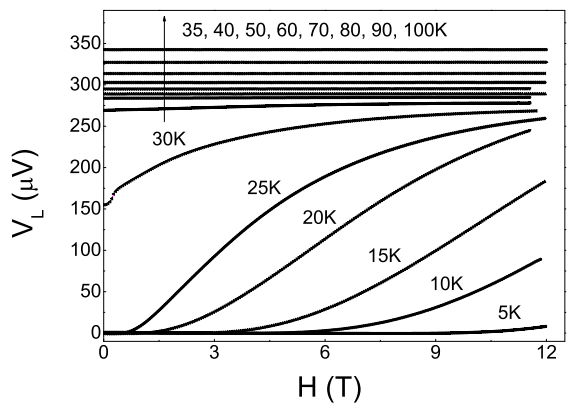} \caption{\label{f5} $H$
dependence of the Longitudinal voltage of
La$_{1.89}$Sr$_{0.11}$CuO$_{4}$ single crystal between the points
C and D with a current of 3 mA between the points of A and B and
without thermal gradient in the configuration shown in FIG.
~\ref{f2} }

\end{figure}

\begin{figure}
\includegraphics [width=.99\columnwidth] {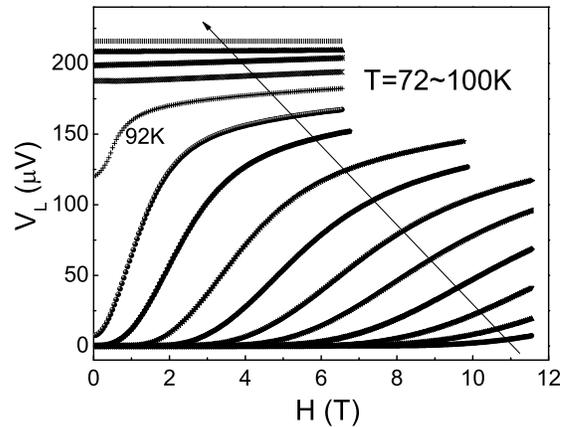}
\caption{\label{f6}$H$ dependence of the Longitudinal voltage of
YBa$_{2}$Cu$_{3}$O$_{7 - \delta }$ thin film between the points C
and D with a current of 3 mA between the points of A and B and
without thermal gradient in the configuration shown in
FIG.~\ref{f2}. The step of temperature is 2 K.}
\end{figure}

The dissipation shown above may be comprised of the contributions
of both the flux flow and the quasiparticle scattering. The new
configuration in our experiment can reduce the dissipative
contribution from the quasiparticle scattering and enhance the
signal due to the possible vortex motion if we send a current
between A-B and measure the voltage between C-D. As shown in FIG.
\ref{f2}, the current was applied in A-B direction, and the
longitudinal voltage between the points on the side of the sample
C and D was measured. As the C-D direction is perpendicular to the
current direction and the two points are located outside the
regime where the main current can reach, so the voltage induced by
the motion of quasiparticles is practically reduced; while the
motion of the vortex crossing C-D induced by the current pump can
be detected and are enhanced relatively as a consequence. The
results of the longitudinal voltage measurement for underdoped
single crystal at the temperature 5$\sim $100 K and optimally
doped YBa$_{2}$Cu$_{3}$O$_{7 - \delta }$ thin film at the
temperature 72$\sim $100 K are shown in FIG. \ref{f5} and FIG.
\ref{f6} respectively. We can see in both figures the similar
dissipation manner as that of the resistance of the samples in
magnetic field shown in FIG. \ref{f4}. The dissipation deriving
from the motion of Abrikosov vortices disappears when the
temperature is above $T_{c}$, where the dissipation of the samples
is independent on magnetic field at fixed temperatures.

%%%%%%%%%%%%%%%
%%%%%%%%%%%%%%%%%%%%%%%%%%%%%%%

The results from the measurement of Nernst voltage, resistance and
longitudinal voltage of our samples at different temperatures may
indicate different vortices below and above $T_{c}$, provided that
the dissipation is attributed to the vortex motion. Apparently,
below $T_{c}$ the dissipation is induced by the motion of the
Abrikosov vortices. The Abrikosov vortex dissipation leads to the
resistance in mixed state and, at the same time, the phase slip
caused by the motion of Abrikosov vortices along the thermal
gradient leads to strong Nernst signal. The longitudinal voltage
below $T_{c}$ shows a non-linear field dependence which indicates
the feature of Abrikosov flux flow.

%%%%%%%%%%%%%%%%%%%%%%%%%%%%%%
%%%%%%%%%%%%%%%

Above $T_{c}$, the dissipation may be caused by the motion of the
spontaneously generated vortex-antivortex pairs. According to the
theory derived by Berezinsky\cite{Berezinsky} and by Kosterlitz
and Thouless\cite{KT}(BKT), just below a transition temperature
$T_{BKT}$ which lies very nearby $T_{c}$, in samples there are
thermally excited vortices and antivortices binding in pairs, even
in the absense of an external magnetic field. Above $T_{BKT}$ the
pairs unbind to form vortex plasma which can flow freely as
pancake vortices with their cores confined inside individual
superconducting layers. The possibility of this BKT transition was
verified and studied in layered HTS
system\cite{Tomoko,Culbertson,Matsuda,Pradhan,Miu,Martin}. In the
region above $T_{c}$, the magnetic field applied on the samples
influences the neutral vortices plasma, and polarizes part of the
vortices. The dissipation caused by vortex and antivortex on the
tolal resistance is the same, although the moving directions of
them are different. In other words, the dissipation resulted from
the current driven vortex motion is not related to the
polarization of, but to the total number of the vortex and the
antivortex in the sample, which is constant when changing $H$. So
the resistance of the sample in magnetic field above $T_{c}$ is
weakly dependent on the magnetic field as illustrated in FIG.
\ref{f4}. For the same reason, the longitudinal voltage induced by
the motion of the vortex plasma in the normal state is also
independent on the magnetic field.

In the case of Nernst effect a different consequence appears in
the pseudogap region. The vortex plasma redirected by the external
magnetic field moves across E-F along the thermal gradient,
leading to the phase slips and a transverse Nernst electric field
$E_{y}$ appears consequently. The direction of the $E_{y}$ caused
by the vortex is different from that of the $E_{y}$ caused by the
antivortex moving along the same direction. So the direction and
strength of $E_{y}$ is not determined by the total number but by
the amount of the difference between vortex and antivortex, i.e.
by the sign and the amount of the net vortex. The density of the
net vortex in the samples increases with increasing external
magnetic field, leading to the increase of Nernst voltage with $H$
in the field range of our experiment. The 2D feature of the Nernst
effect in the pseudogap region of underdoped cuprate
superconductors has been proved by the experiments of Wen
\textit{et al}\cite{Wen}, which supports our postulation above.

\subsection{Manipulating Vortex Motion by Thermal and Lorentz Force}

%%%%%%%%%%%%%%%%%%%%%%%%%%%%%%
\begin{figure}[t]
\includegraphics [width=.99\columnwidth] {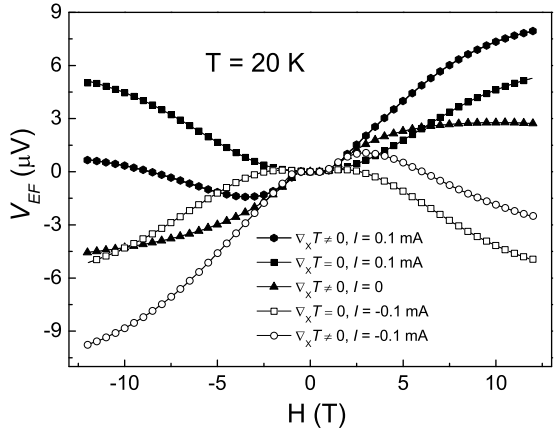}
\caption{\label{f9}The result of $V_{EF}$ at the temperature of 20
K with/without a temperature gradient along the length direction
and a transverse current along A-B on the
La$_{1.89}$Sr$_{0.11}$CuO$_{4}$ single crystal. Different symbols
represent the case of $I$ = 0 (solid triangles), $I$ = 0.1 mA
(solid circles), $I$ = -0.1 mA (open circles) with thermal
gradient, and $I$ = 0.1 mA (solid squares), $I$ = -0.1 mA (open
squares) without thermal gradient.The heating power is 1.0 mW. For
the convenience of comparison, all the curves have been shifted
along $V_{EF}$ axis and the values of $V_{EF}$ is zero when $H$ =
0 after the shift.}
\end{figure}

%%%%%%%%%%%%%%%

The Nernst signal and the longitudinal voltage as the result from
the motion of vortex manipulated solely by the thermal gradient
and solely by Lorentz force are discussed above. Now let's go on
to discuss the result of manipulating the motion of the vortex by
the thermal gradient and the Lorentz force together. With a
transverse current applied along A-B direction and a temperature
gradient along the length direction of the
La$_{1.89}$Sr$_{0.11}$CuO$_{4}$ single crystal at the same time,
the voltage $V_{EF}$ between the Nernst leads E and F are
measured.

%%%%%%%%%%%%%%%%%%%%%%%%%%%%%%%%%%%%%%%%%%%%%%%%%%%%%%%%%%%%%%%%%
\begin{figure}
\includegraphics [width=.99\columnwidth] {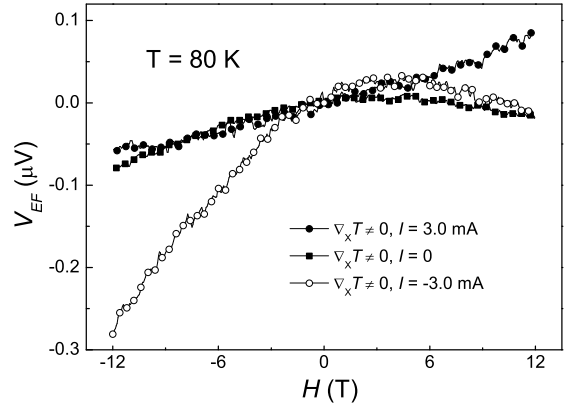}
\caption{\label{f7}The result of $V_{EF}$ at the temperature of 80
K with a temperature gradient along the length direction and a
transverse current along A-B on the
La$_{1.89}$Sr$_{0.11}$CuO$_{4}$ single crystal. The solid squares,
solid circles and oper circles present the case of 0 mA, 3.0 mA
and -3.0 mA respectively. The heating power is 1.0 mW. All the
curves have been vertically shifted to make the data comparable as
that in FIG.~\ref{f9}.}
\end{figure}
%%%%%%%%%%%%%%%%%%%%%%%%%%%%%%%%%%%%%%%%%%%%%%%%%%%%%%%%%%%%%%%%%%%

FIG.~\ref{f9} shows the result of $V_{EF}$ at 20 K for the case
$I$ = 0, $I$ = 0.1 mA, and $I$ = -0.1 mA with and without thermal
gradient. $V_{EF}$ with thermal gradient and without current
(presented by solid triangles) is the Nernst signal at this
temperature, where we can see the sign of $V_{EF}$ is dependent on
the direction of magnetic field. For the case of $\nabla$$_{x} T $
= 0 and $I \neq$ 0, $V_{EF}$ (solid and open squares) shows the
characteristics of Abrikosov vortices and the sign of the $V_{EF}$
after vortices melting is independent on the direction of current.
The features of the curves is similar to those in FIG.~\ref{f4}.
The results manipulating vortex motion by thermal and Lorentz
force is shown in the case of $\nabla$$_{x} T \neq$ 0 and $I$ =
$\pm$ 0.1 mA (solid and open circles) in FIG.~\ref{f9}. All the
curves show that the transport properties in mixed state is
dominated by Abrikosov vortex motion.

FIG.~\ref{f7} shows the result at 80 K with the current of 0 mA, 3
mA and -3 mA. In FIG.~\ref{f7}, all the curves have been moved
along $V_{EF}$ axis and hence the values of $V_{EF}$ is zero when
$H$ = 0 for the convenience of comparison. The influence of the
Faraday effects and the resistive components of the data are taken
away by moving the curves. We can see from FIG.~\ref{f7} that with
positive transverse current, the $V_{EF}$ measured under negative
magnetic field is almost equivalent to the $V_{EF}$ without
current, but for the case under positive field, there is an
obvious difference between the $V_{EF}$ with current and that
without current. With negative transverse current, the $V_{EF}$
measured under positive magnetic field is almost equivalent to the
$V_{EF}$ without current, but for the case under negative field,
there is an obvious difference between the $V_{EF}$ with current
and that without current. So we can conclude that the effects of
the Lorentz force on the Nernst effect in the normal state is
related to the direction of the magnetic field. This phenomenon
can not be explained by the effects of quasiparticles and may be
explained only by the motion of vortex. So the Nernst signal above
$T_{c}$ is also contributed by flux motion, although the vortex
structure and feature may be very different below and above
$T_{c}$.

\section{Conclusions}

The in-plane resistance, Nernst voltage and longitudinal voltage
in a new experimental configuration of the underdoped
La$_{1.89}$Sr$_{0.11}$CuO$_{4}$ single crystal and optimally doped
YBa$_{2}$Cu$_{3}$O$_{7 - \delta }$ thin film are measured by
sweeping magnetic field at fixed temperatures. The vortex motion
is successfully manipulated by using thermal and / or Lorentz
force in the mixed and normal states of the samples. The Nernst
signal in the normal state is observed. The signal maybe caused by
the motion of quasiparticles or vortex or both. Our results
indicate that Nernst signal above $T_{c}$ may be contributed by
vortex motion, though the contribution of the quasiparticle has
not been excluded. Different vortices below and above $T_{c}$ are
expected if the strong Nernst signal in the pseudogap region was
attributed to the vortex motion. Below $T_{c}$ the dissipation is
induced by the motion of the Abrikosov vortices. Above $T_{c}$ the
dissipation may be caused partly by the motion of the
spontaneously generated unbinded vortex-antivortex pairs (vortices
plasma).

\section{Acknowledgements}

This work is supported by the National Natural Science Foundation
of China (NSFC), the Ministry of Science and Technology of China,
and the Chinese Academy of Sciences.

\end{document}